# Ordered Weighted Average Based grouping of nanomaterials with Arsinh and Dose Response similarity models

Alex Zabeo[1*], Gianpietro Basei[1], Georgia Tsiliki[2], Willie Peijnenburg[3,4], Danail Hristozov[1*]

[1] GreenDecision Srl., Venezia, Italy

[2] Athena RC, Athens, Greece

[3] National Institute of Public Health and the Environment (RIVM), Center for Safety of Substances and Products, Bilthoven, The Netherlands

[4] Leiden University, Institute of Environmental Sciences (CML), P.O. Box 9518, 2300 RA Leiden, The Netherlands

*Corresponding authors

- **Abstract:** In the context of the EU GRACIOUS project, we propose a novel procedure for similarity assessment and grouping of nanomaterials. This methodology is based on the (1) Arsinh transformation function for scalar properties, (2) full curve shape comparison by application of a modified Kolmogorov–Smirnov metric for bivariate properties, (3) Ordered Weighted Average (OWA) aggregation-based grouping distance, and (4) hierarchical clustering. The approach allows for grouping of nanomaterials that is not affected by the dataset, so that group membership will not change when new candidates are included in the set of assessed materials. To facilitate the application of the proposed methodology, a software script was developed by using the R programming language which is currently under migration to a web tool. The presented approach was tested against a dataset, derived from literature review, related to immobilisation of *Daphnia magna* and reporting information on several nanomaterials and properties.

## 1 Introduction

The large diversity of nanoforms (NFs) used in nano-enabled products has made their case-by-case safety assessment very demanding in terms of resources. It has been widely accepted by regulators, industries, and scientists that the implementation of robust approaches for similarity assessment as a basis for grouping could help to optimise testing costs and the use of experimental animals (ECHA 2019a; 2019b). The grouping of similar NFs can enable read across of essential information for both safe by design and regulatory risk assessment purposes. To facilitate this, the European Commission funded Horizon 2020 GRACIOUS project (https://www.h2020gracious.eu), which has developed a framework to guide stakeholders from industry, consultancies and regulation in the process of grouping NFs (Stone et al. 2020).

To support the grouping, GRACIOUS has also proposed an array of methods to assess similarity between NFs in terms of intrinsic and extrinsic physicochemical characteristics as well as toxicity, either via a pairwise analysis conducted property-by-property, or by assessing all relevant properties and hazard endpoints simultaneously via multidimensional analysis (Jeliazkova et al. 2021). Such methods are based for example on an x-fold comparison (Janer, Landsiedel, and Wohlleben, 2021), Euclidean distance, Bayesian logic (Tsiliki et al. 2021) or clustering methods (Jeliazkova et al. 2021).

These statistical approaches apply different algorithms to compare NFs, but what unifies them is that in all of them, the grouping is relative to the dataset. This means that changes to the dataset due to adding information on new materials might theoretically cause some of the NFs in the initial dataset to change their group membership. This can have important implications on any read-across and/or risk assessment studies already performed and on the respective risk management and/or regulatory decisions made. The likelihood that this theoretical possibility becomes real is particularly large for sectors where new NFs of the same of similar substances are continuously being developed (e.g. organic and inorganic pigments, silicas, carbon nanotubes), and new health and safety information that is relevant to include in grouping datasets is constantly emerging.

To address this issue, the objective of this paper is to propose a methodology for assessment of similarity between NFs, which enables grouping of the NFs that is not affected by the dataset. This will guarantee that the group membership of the NFs will not change when new candidates are included in the set of assessed materials. This approach is based on a combination of the (1) Arsinh transformation function for scalar properties, (2) full curve shape comparison by application of a modified Kolmogorov–Smirnov metric for bivariate properties, (3) Ordered Weighted Average (OWA) aggregation-based grouping distance, and (4) hierarchical clustering. Specifically, the methodology first applies the Arsinh transformation to the distance between two NFs, and then rescales the result to the Arsinh of a biologically relevant threshold, such as a multiple of the positive control value for the same property. This metric distance-based similarity allows the final aggregated distance between NFs to not be affected by the dataset, preserving symmetry and triangular inequality, leading to groups which do not change if new members are included in the assessment. The rescaled similarity matrices are utilized for grouping by applying agglomerative hierarchical clustering in a multidimensional space. To evaluate the multidimensional distance, OWA aggregation (Yager and Kacprzyk 1997) is applied, where the highest distances among all dimensions are aggregated as the overall NF distance from other NFs.

To facilitate the application of the proposed methodology a software script was developed by using the R programming language. The script is currently under migration into a web application. The presented approach was tested against a dataset, derived from a literature review. The dataset is related to immobilisation of *Daphnia magna* and includes information on several nanomaterials and properties.

## 2 Methods

The aim of the proposed methodology is to perform grouping of NFs considering all the available data such as physicochemical and toxicological information. The idea is that by integrating the different data types and applying specific transformations to each of them it is possible to achieve threshold-based grouping that is not affected by the dataset.

The main feature of the proposed methodology is the application of metric distance-based similarity among the different considered parameters so that the final aggregated distance between NFs: i) would not be affected by the dataset: ii) preserves symmetry; and iii) ensures triangular inequality. This leads to groups which do not change if new members are included in the assessment.

Given a set of candidate NFs characterised by several intrinsic and extrinsic properties, the methodology elicits possible groups of 'sufficiently similar' NFs. To this end the following phases are foreseen:

- Both scalar and dose response data is transformed and scaled to become comparable.
- Single properties' distances are calculated with specific methods for scalar and dose response data.
- Pairwise NFs distance is calculated for all possible pairs of candidates by considering all properties.
- Hierarchical clustering is performed using the pairwise distances.

To aid the application of the proposed methodology, a software script programmed in the R programming language was developed. The script reads csv input tables containing properties data for each candidate NF, performs transformation and scaling, evaluates pairwise distances, and applies hierarchical clustering. Finally, it provides several charts and intermediate results for all performed calculations while suggesting the final inferred NFs groups. Moreover, a development effort is currently undergoing to migrate the Script into an online tool.

### 2.1 Data transformation and scaling

The first step of the procedure involves initial data treatment. Only scalar data is transformed to become manageable, reduce the impact of errors and make the different properties comparable.

The proposed transformation is the Arsinh function (i.e., Inverse hyperbolic sine, Figure 1). The Arsinh function has similar characteristics to the logarithm function (upon which it is based) but presents several comparative advantages: it is continuous and defined over the whole real numbers' domain (including negative numbers), it is always non-negative for positive real numbers (including the [0,1] portion) and is less steep than the logarithm function for small values in (0,1) which helps to better distinguish upon small numbers.

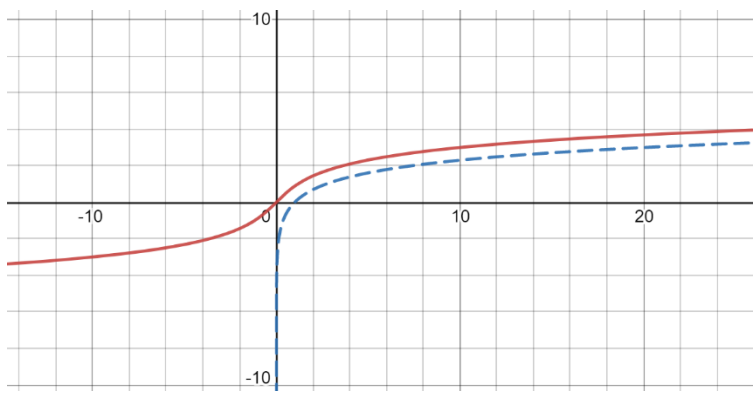

**Figure 1:** Arsinh (red line) vs ln (blue dash) comparison.

After being transformed, data are scaled with the aim of normalizing the distances upon different properties values. This allows for a meaningful integration when calculating a multidimensional distance before clustering (see 2.4). One of the aims of the proposed methodology is its ability of maintaining the same groups even if new members are included in the study, i.e. being an absolute rather than relative assessment. To preserve this feature, scaling cannot be based on statistical descriptors as usually done in such situations (e.g. by evaluating the standardized values). Instead, a specific threshold is set for each property as scaling ratio. This means that the method provides absolute results given that scaling thresholds are not changed between compared applications. Such a threshold should be property specific, although data agnostic, and should represent the biological distance sufficient to define two items as not similar. In the developed R script such threshold has been established empirically from the GRACIOUS data as a multiple of negative control for several widely used properties. Such negative controls are derived from benchmark or reference materials used in GRACIOUS, all of which are well-characterised nanomaterials, e.g. from the JRC repository. Such selection respects the absolute result prerequisite, given that the negative control is kept fixed for the same property in different assessments.

## 2.2 Evaluation of distances of single properties

To evaluate the distance of scalar properties the simple single dimension Euclidean distance is used (i.e. $|x_1 - x_2|$) whereas for dose response (i.e. bivariate) data the methodology is based on full dose response comparison.

The basic idea is that considering the entire dose response data points can help to better assess similarity than only considering a single element of the dose response relationship such as the point of departure (e.g., BMDL, NOAEL) or another reference dose (e.g. LC50). This is because different NFs could have similar NOAEL or LC50s, but very different slopes and shapes of their overall dose response curves.

Measured concentrations of experiments conducted for different NFs are in general not coincident both in numerosity and value. This implies that measuring effects distances upon the empirical measured data would in general be unfeasible as it would not be possible to associate the different responses to be compared to a representative concentration. Because of this, the proposed methodology is not based upon the empirical data points but rather their statistically fitted curves. To fit the curves the PROAST model (Slob 2002) from RIVM is applied. PROAST is suitable for statistically sampling both *in vitro* and *in-vivo* dose-response data as it includes several kinds of fitting distributions.

Once all curves are fitted, their distance is assessed by the Jensen–Shannon distance. The Jensen–Shannon distance is defined as the square root of the Jensen–Shannon divergence (Lin 1991) and has the advantage, upon the latter, of being a proper metric (i.e. a distance function that respects identity, symmetry and triangular inequality). The Jensen–Shannon divergence is a well-known method to measure similarity between two probability distributions, it is based on the Kullback–Leibler divergence (Kullback and Leibler 1951), with the enhancements of being symmetric and always having a finite value.

Once pairwise distances between all NFs have been evaluated both for scalar and dose response data, the result is a three-dimensional distance matrix reporting distance between all NFs among all properties which can be used in any multidimensional distances-based similarity algorithm.

## 2.3 Evaluation of a single distance value integrating all properties' distances

To apply any clustering technique, it first is necessary to integrate distances among all single properties into an overall distance value, generating a bidimensional distance matrix of pairwise NF distances. Several distance metrics are possible candidates for this step of the process, e.g. Euclidean distance, Manhattan distance, etc. All the aforementioned distances are based on topographical concepts of physical distance between objects. As in this setting we are assessing dissimilarity rather than distance, we decided to apply an aggregation function coming from Multi Criteria Decision Analysis (MCDA) (Zabeo et al. 2011; Giove et al. 2009) and more specifically from its Multi Attribute Value Theory (MAVT) (Angelis and Kanavos 2017; D. R. Hristozov et al. 2014) branch. While both MCDA and MAVT aim at ranking different alternatives based on selected characteristics, the basic underlying aggregation functions used in MAVT are deemed to properly resemble the kind of dissimilarity metric which fits the NF grouping issue. More specifically, the proposed methodology makes use of the Ordered Weighted Average (OWA) aggregation operator (D. Hristozov et al. 2016; Ahn and Yager 2014) which is a generalization of the minimum, average and maximum operators which are also the foundation of the aforementioned topographical metrics (e.g. Euclidean distance is based on the average aggregation function).

The OWA operator is based on a set of weights which are used as operational parameters to adapt the general OWA formula to a specific implementation, in fact the OWA operator of dimension $n$ is a mapping function characterized by a weights vector $w = (w_1, \ldots, w_n)$, where $w_i \in [0,1]$ and $\sum_{i=1}^{n} w_i = 1$, and defined as:

$$\text{OWA}(x_1, \ldots, x_n) = \sum_{i=1}^{n} w_i \cdot b_i$$

Where $b_i$ is the $i$th largest element in the $(x_1, \ldots, x_n)$ vector, so that $b_1 \geq b_2 \geq \cdots \geq b_n$.

Given that for a generic multidimensional grouping problem the number of properties to aggregate in a single distance score should always be greater or equal to three, the proposed generic weights vector is $w = (0.7, 0.2, 0.1, 0 \ldots)$ where all weights from fourth to $n$th (if $n > 3$) are set to 0. The proposed weights were defined empirically following the idea that similarity should not be based on compensatory aggregation but rather that a relevant distance in just a few properties should determine group separation.

### 2.4 Clustering and definition of similar groups

The final step of the proposed grouping methodology consists in applying agglomerative hierarchical clustering (Ah-Pine 2018) to the pairwise distance matrix obtained in the previous step. The agglomerative clustering process starts by having each element (i.e. each NF) in a separate cluster, then, utilizing the complete-linkage protocol which combines two clusters basing on the farthest pair of their elements, clusters are joined together hierarchically, forming a tree structure usually visualized as a dendrogram. The complete-linkage protocol was selected so that the cutting threshold of 1 can successively be used to establish final groups while maintaining the immutability of groups when new members are included in the assessment.

The obtained tree of growing clusters is finally used to suggest possible groups of sufficiently similar NFs. As the selected rescaling threshold (see 2.1) was based on significant biological difference, it is now possible to use 1 as cutting threshold for the group's boundaries, so that all clusters more than a unit apart represent sufficiently dissimilar groups. This can easily be represented in a dendrogram where a horizontal line at distance 1 is drawn.

### 3 Results

The presented methodology has been tested by its application to a literature review-based dataset related to immobilization of *Daphnia magna* exposed to five different NFs of Nano Copper Oxide. The dataset contains immobilization dose-response data. Two different parameters were used for multidimensional similarity: Primary size diameter (scalar number) measured by Dynamic Light Scattering and immobilization (dose-response dataset). The complete dataset is available in the supplementary material, the five assessed NFs are derived from the following papers:

- Kim_40: (Kim et al. 2017) 40 nm data
- Santos-Rasera_40: (Santos-Rasera et al. 2019) 40 nm data

- Santos-Rasera_80: (Santos-Rasera et al. 2019) 80 nm data
- Seo_40: (Seo et al. 2014) 40 nm data
- Sovova_50: (Sovova, Kočí, and Kochankova 2009) 50 nm data

The results presented below should be considered as a proof-of-concept aimed at testing the proposed methodology on real data. This should therefore not be considered a complete study providing reliable regulatory level results as this is out of the scope of this manuscript.

### 3.1 Data transformation and scaling

Transformation and scaling are only applied to scalar properties as dose-response distances are managed separately. In this application size data has been transformed, by Arsinh, and scaled. The scaling threshold was established empirically from the GRACIOUS data as a multiple of negative control and is the proposed default scaling factor for size in general applications.

### 3.2 Evaluation of single properties' distances

Distances among properties are calculated through different procedures according to their type, while for scalar properties Euclidean distance is used. For dose-response data the process involves curve fitting and distance evaluation as explained before.

In Figure 2 the original unfitted data is presented while the corresponding fitted curves are in Figure 3.

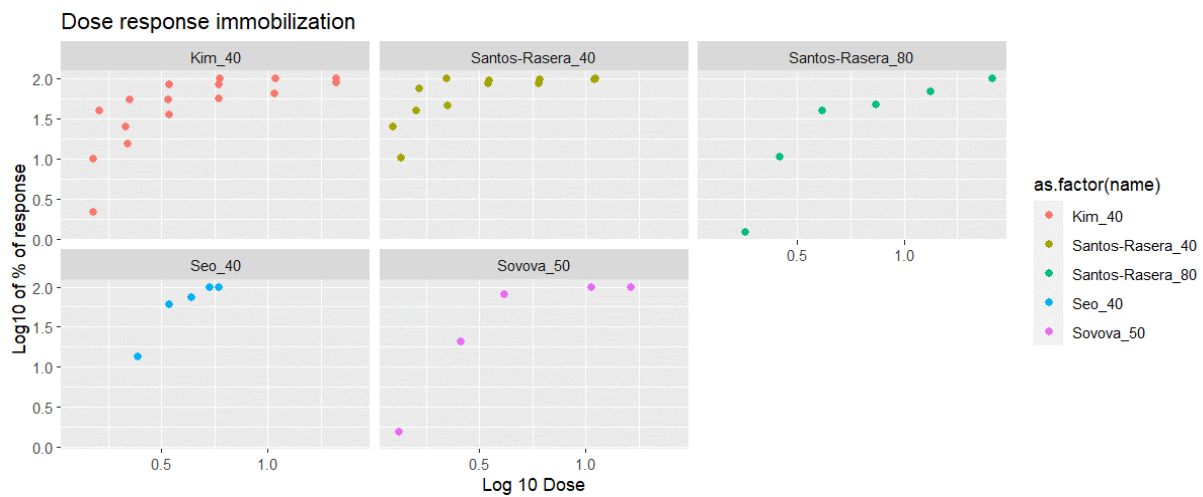

**Figure 2:** Original dose response data

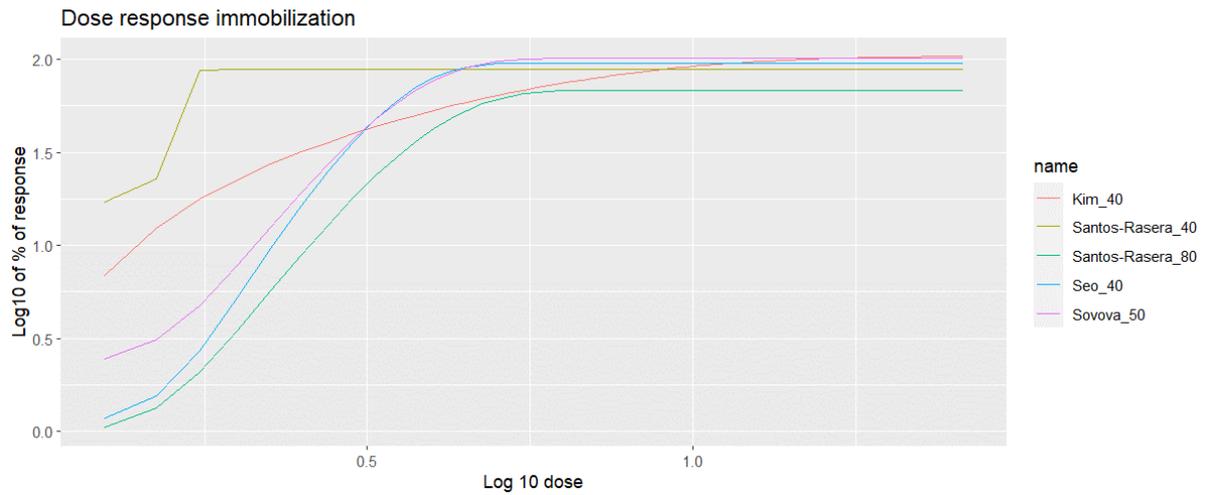

**Figure 3:** Dose response fitted curves using the PROAST software

The fitted curves all present similar shapes even though some notable differences are present. All fitted curves are based on the Exponential function family but, while *Kim_40* is fitted by a 4 parameters' version, all others are fitted by the complete version with 5 parameters. Moreover, among the 5 parameters fitted NFs, *Santos-Rasera_40* is the one with much different parameters' values compared to the others (see SI.2).

Distance between the fitted lines has been calculated by first rescaling all the curves to the [0,1] domain using overall minimum and maximum values as boundaries and then by square rooting their Jensen–Shannon divergence. The obtained distances both for the Size scalar property and Immobilization dose response property are reported in Figure 4 below. It is already visible that *Santos-Rasera_80* presents higher distances in general.

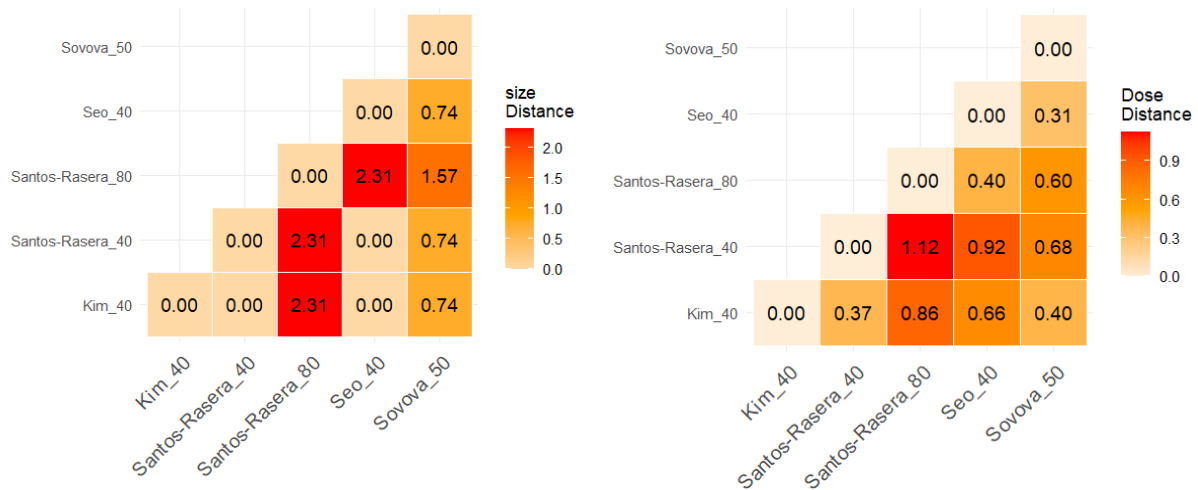

**Figure 4:** Pairwise distances for Size distribution and Dose response, each cell shows the pairwise distance among NFs according to Size (left) and immobilization dose response (right).

## 3.3 Evaluation of a single distance value integrating all properties' distances

The OWA aggregation function with default weights was applied to aggregate the two previously obtained single property distances into an overall distance matrix presented in Figure 5. As previously deemed *Santos-Rasera_80* still presents the highest distances overall.

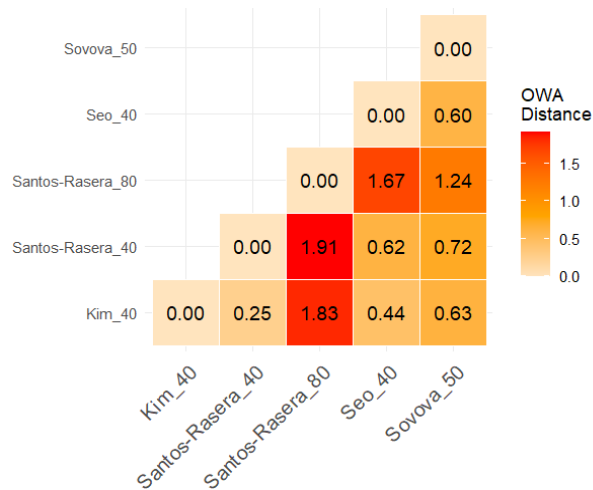

**Figure 5:** Aggregated distance evaluation through the OWA operator with default weights, each cell shows the pairwise distance among NFs according to distances calculated through the OWA operator which aggregates the size and immobilization distances calculated in the previous step.

## 3.4 Clustering and definition of similar groups

The final step of assessment involves the application of standard hierarchical clustering based on the OWA weights generated in the previous step. The result of the clustering process is the dendrogram presented in Figure 6. As explained before, different groups can be formed by cutting the obtained dendrogram at distance equal to 1, which in this example generates two groups, one formed by *Santos-Rasera_80* alone and the other containing all other NFs.

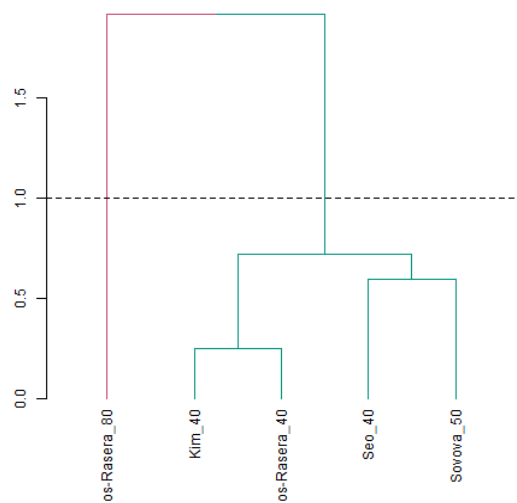

**Figure 6:** Final clustering dendrogram generating two groups expressed by red and green lines. Each cluster (represented by a branch) is vertically positioned corresponding to the distance between its elements, *Santos-Rasera_80* forms a cluster of its own and is separated from the cluster of all other NFs as its distance is above one, as explained in 2.4.

## 4 Discussion

In this manuscript we proposed a novel procedure for similarity assessment and grouping of NFs. The proposed methodology accomplishes the main objective of this work, by providing a static similarity and grouping approach which does not change as new NFs are included in the assessment. The main limitation of this approach is related to the need of scaling thresholds for the different scalar properties. These thresholds are used to rescale different properties, each with its own unit of measure, to a generic scale where a distance of one unit represents biological relevance. If such thresholds are maintained fixed among several applications of the proposed assessment, then the obtained groups will be coherent. This means that the same NF will always stay in the same group even if other NFs are included in the assessment. To overcome this issue, default thresholds for the most common properties were included in the developed R script, as empirically derived from negative control substances of GRACIOUS experiments. The script also allows using more specific thresholds set by the user.

To confirm the strength of the methodology, it was applied to literature dose response data related to immobilization for different Copper Oxide based NFs. The assessed NFs were characterized by two properties, size diameter, a scalar number and percentage dose response curves for immobilization of *Daphnia magna*. The original data was scaled and transformed to move from the original unit of measure, different for each property, to comparable quantities related to biological relevance so that a distance of one unit has the same meaning for all the different properties. Transformed data was then aggregated into a single distance used to create groups of sufficiently similar NFs according to all the assessed properties. The study showed how *Santos-Rasera_80* should constitute a group per se while all other NFs should be grouped together as having an aggregated distance above one. The procedure correctly separated the NFs with higher differences in the values of both parameters in a transparent procedure.

These results are in line with those from other similarity and grouping methods applied in the GRACIOUS project. However, it is important to note that these results should only be considered as a proof-of-concept aimed at testing the presented approach and the R script with real data. The results of the study presented here should therefore not be considered as providing regulatory level similarity assessment results.

Several future enhancements are foreseen for the proposed methodology and the R script. This will allow to take uncertainty into consideration and mitigate bias related to the comparison of fitted dose-response curves generated from original data points with non-intersecting dose ranges.

## 5 Conclusions

We proposed an Arsinh and OWA based methodology for similarity assessment and grouping which preserves groups even if new NFs are included in the assessment. The approach involves four stages: (1) data transformation and scaling based on the Arsinh function for scalar properties; (2) evaluation of single properties' distance by Kolmogorov–Smirnov metric based full curve shape comparison for bivariate properties; (3) Evaluation of a single integrated distance using Ordered Weighted Average (OWA) aggregation; and (4) clustering and definition of similar groups with hierarchical clustering. To facilitate the application of this methodology, an R script was developed, which is currently undergoing migration into a web-based tool.

Positive aspects of the proposed approach are mainly related to its absolute grouping (as opposed to relative grouping) where group membership is not changing when new candidates are included in the set of assessed materials. This feature is highly relevant for nanomaterials as in practice there are often many existing and emerging nanoforms of the same substance (e.g., pigments) that may not be included in the initial dataset but may be added to the grouping as they are actually developed and information for them becomes available. On the other hand, the proposed method requires parameter specific thresholds which could be difficult to establish for some parameters. To demonstrate the methodology and the R script, they were tested in a proof-of-concept exercise against a literature-based dataset for immobilization. The method was also validated against other methods in (Jeliazkova et al. 2021) in this same issue, results were also consistent with conclusions from other methods applied to the same data in the GRACIOUS project (Tsiliki et al. 2021). This confirmed the validity and the soundness of the proposed approach.

**Acknowledgements**

The GRACIOUS project has received funding from the European Union's Horizon 2020 research and innovation programme under grant agreement No 760840.

1 **Supplementary Information**

2 **SI.1 Original dataset**

| type | category | name | article | producer | Size (nm) | Dose (mg/L) | Response (%) |
|---|---|---|---|---|---|---|---|
| DLS | nCuO | Kim_40 | Kim, S., et al. (2017) | Alfa Aesar (Ward Hill, USA) | 40 | 0.63 | 39.63 |
| DLS | nCuO | Kim_40 | Kim, S., et al. (2017) | Alfa Aesar (Ward Hill, USA) | 40 | 1.25 | 55.37 |
| DLS | nCuO | Kim_40 | Kim, S., et al. (2017) | Alfa Aesar (Ward Hill, USA) | 40 | 2.45 | 84.65 |
| DLS | nCuO | Kim_40 | Kim, S., et al. (2017) | Alfa Aesar (Ward Hill, USA) | 40 | 4.95 | 99.63 |
| DLS | nCuO | Kim_40 | Kim, S., et al. (2017) | Alfa Aesar (Ward Hill, USA) | 40 | 9.95 | 100.00 |
| DLS | nCuO | Kim_40 | Kim, S., et al. (2017) | Alfa Aesar (Ward Hill, USA) | 40 | 20.05 | 100.00 |
| DLS | nCuO | Kim_40 | Kim, S., et al. (2017) | Alfa Aesar (Ward Hill, USA) | 40 | 0.52 | 10.01 |
| DLS | nCuO | Kim_40 | Kim, S., et al. (2017) | Alfa Aesar (Ward Hill, USA) | 40 | 1.15 | 25.37 |
| DLS | nCuO | Kim_40 | Kim, S., et al. (2017) | Alfa Aesar (Ward Hill, USA) | 40 | 2.40 | 54.65 |
| DLS | nCuO | Kim_40 | Kim, S., et al. (2017) | Alfa Aesar (Ward Hill, USA) | 40 | 4.90 | 84.99 |
| DLS | nCuO | Kim_40 | Kim, S., et al. (2017) | Alfa Aesar (Ward Hill, USA) | 40 | 9.95 | 100.00 |
| DLS | nCuO | Kim_40 | Kim, S., et al. (2017) | Alfa Aesar (Ward Hill, USA) | 40 | 19.95 | 100.00 |
| DLS | nCuO | Kim_40 | Kim, S., et al. (2017) | Alfa Aesar (Ward Hill, USA) | 40 | 0.52 | 2.16 |
| DLS | nCuO | Kim_40 | Kim, S., et al. (2017) | Alfa Aesar (Ward Hill, USA) | 40 | 1.20 | 15.36 |
| DLS | nCuO | Kim_40 | Kim, S., et al. (2017) | Alfa Aesar (Ward Hill, USA) | 40 | 2.45 | 35.73 |
| DLS | nCuO | Kim_40 | Kim, S., et al. (2017) | Alfa Aesar (Ward Hill, USA) | 40 | 4.90 | 55.71 |
| DLS | nCuO | Kim_40 | Kim, S., et al. (2017) | Alfa Aesar (Ward Hill, USA) | 40 | 9.79 | 65.35 |

| DLS | nCuO | Kim_40 | Kim, S., et al. (2017) | Alfa Aesar (Ward Hill, USA) | 40 | 19.95 | 89.66 |
| --- | --- | --- | --- | --- | --- | --- | --- |
| DLS | nCuO | Santos-Rasera_40 | Santos-Rasera, J. R., et al. (2019) | US Nanomaterials Research Inc | 40 | 0.35 | 10.35 |
| DLS | nCuO | Santos-Rasera_40 | Santos-Rasera, J. R., et al. (2019) | US Nanomaterials Research Inc | 40 | 0.60 | 39.63 |
| DLS | nCuO | Santos-Rasera_40 | Santos-Rasera, J. R., et al. (2019) | US Nanomaterials Research Inc | 40 | 1.25 | 45.70 |
| DLS | nCuO | Santos-Rasera_40 | Santos-Rasera, J. R., et al. (2019) | US Nanomaterials Research Inc | 40 | 2.47 | 85.71 |
| DLS | nCuO | Santos-Rasera_40 | Santos-Rasera, J. R., et al. (2019) | US Nanomaterials Research Inc | 40 | 5.00 | 86.06 |
| DLS | nCuO | Santos-Rasera_40 | Santos-Rasera, J. R., et al. (2019) | US Nanomaterials Research Inc | 40 | 10.08 | 100.00 |
| DLS | nCuO | Santos-Rasera_40 | Santos-Rasera, J. R., et al. (2019) | US Nanomaterials Research Inc | 40 | 0.25 | 25.37 |
| DLS | nCuO | Santos-Rasera_40 | Santos-Rasera, J. R., et al. (2019) | US Nanomaterials Research Inc | 40 | 0.65 | 74.98 |
| DLS | nCuO | Santos-Rasera_40 | Santos-Rasera, J. R., et al. (2019) | US Nanomaterials Research Inc | 40 | 1.22 | 98.91 |
| DLS | nCuO | Santos-Rasera_40 | Santos-Rasera, J. R., et al. (2019) | US Nanomaterials Research Inc | 40 | 2.52 | 95.01 |
| DLS | nCuO | Santos-Rasera_40 | Santos-Rasera, J. R., et al. (2019) | US Nanomaterials Research Inc | 40 | 5.07 | 98.57 |
| DLS | nCuO | Santos-Rasera_40 | Santos-Rasera, J. R., et al. (2019) | US Nanomaterials Research Inc | 40 | 10.00 | 98.23 |
| DLS | nCuO | Santos-Rasera_80 | Santos-Rasera, J. R., et al. (2019) | US Nanomaterials Research Inc | 80 | 0.79 | 1.20 |
| DLS | nCuO | Santos-Rasera_80 | Santos-Rasera, J. R., et al. (2019) | US Nanomaterials Research Inc | 80 | 1.61 | 10.71 |
| DLS | nCuO | Santos-Rasera_80 | Santos-Rasera, J. R., et al. (2019) | US Nanomaterials Research Inc | 80 | 3.16 | 40.17 |
| DLS | nCuO | Santos-Rasera_80 | Santos-Rasera, J. R., et al. (2019) | US Nanomaterials Research Inc | 80 | 6.40 | 47.03 |
| DLS | nCuO | Santos-Rasera_80 | Santos-Rasera, J. R., et al. (2019) | US Nanomaterials Research Inc | 80 | 12.39 | 67.86 |
| DLS | nCuO | Santos-Rasera_80 | Santos-Rasera, J. R., et al. (2019) | US Nanomaterials Research Inc | 80 | 25.06 | 98.80 |
| DLS | nCuO | Seo_40 | Seo, J., et al. (2014) | Alfa Aesar (Ward Hill, USA) | 40 | 1.46 | 13.57 |
| DLS | nCuO | Seo_40 | Seo, J., et al. (2014) | Alfa Aesar (Ward Hill, USA) | 40 | 2.43 | 60.27 |
| DLS | nCuO | Seo_40 | Seo, J., et al. (2014) | Alfa Aesar (Ward Hill, USA) | 40 | 3.41 | 73.50 |
| DLS | nCuO | Seo_40 | Seo, J., et al. (2014) | Alfa Aesar (Ward Hill, USA) | 40 | 4.34 | 100.00 |
| DLS | nCuO | Seo_40 | Seo, J., et al. (2014) | Alfa Aesar (Ward Hill, USA) | 40 | 4.92 | 100.00 |
| DLS | nCuO | Sovova_50 | Sovová, T., et al. (2009) | Aldrich | 50 | 0.32 | 1.57 |
| DLS | nCuO | Sovova_50 | Sovová, T., et al. (2009) | Aldrich | 50 | 1.58 | 20.50 |
| DLS | nCuO | Sovova_50 | Sovová, T., et al. (2009) | Aldrich | 50 | 3.15 | 80.14 |
| DLS | nCuO | Sovova_50 | Sovová, T., et al. (2009) | Aldrich | 50 | 9.64 | 100.00 |
| DLS | nCuO | Sovova_50 | Sovová, T., et al. (2009) | Aldrich | 50 | 15.35 | 100.00 |

**SI.2 Fitted curves**

| Name | Fitting function | a | b | c | d |
|---|---|---|---|---|---|
| Kim_40 | E4-CED: y = a * [c-(c-1)exp(-bx)] | 0.0216 | 0.2357 | 4756.92105 | NA |
| Santos-Rasera_40 | E5-CED: y = a * [c-(c-1)exp(-bx^d)] | 16.06146 | 85.26 | 5.400902 | 10 |
| Santos-Rasera_80 | E5-CED: y = a * [c-(c-1)exp(-bx^d)] | 0.012 | 0.03736 | 5596.894127 | 2.944149 |
| Seo_40 | E5-CED: y = a * [c-(c-1)exp(-bx^d)] | 0.13568 | 0.04458 | 696.394908 | 3.346014 |
| Sovova_50 | E5-CED: y = a * [c-(c-1)exp(-bx^d)] | 1.371939 | 0.05675 | 72.889555 | 2.913248 |